\documentclass[prd,showpacs,nofootinbib,preprintnumbers,
preprint
]{revtex4}
\usepackage{amsmath}  \usepackage{amsfonts}
\usepackage{graphicx,
xcolor
}
\def\be{\begin{equation}}
\def\ee{\end{equation}}
\def\ba{\begin{eqnarray}}
\def\ea{\end{eqnarray}}

\def\lb{\label}

\def\la{\lambda}

\def\tph{{\hat\varphi}}
\def\lb{\label}
\def\W{W}
\def\X{{\hat g}}
\def\R{{\hat R}}
\def\V{{\hat V}}
\def\N{{\hat N}}
\def\a{{\hat a}}

\def\s{{\hat s}}
\def\hnabla{\hat{\nabla}}
\def\hGamma{\hat{\Gamma}}
\def\fl{\mu}
\def\fL{\lambda}

\def\hBox{{\hat \Box}}

\begin{document}
\title{
Conformal and kinetic couplings as two Jordan frames of the same theory}
\author{Dmitri Gal'tsov} \email{galtsov@phys.msu.ru}
\affiliation{Faculty of Physics,
Moscow State University, 119899, Moscow, Russia}
 
\begin{abstract}
Non-minimal scalar-tensor (ST) theories may admit an Einstein frame representation, where gravity is described by the Einstein-Hilbert action plus the scalar sector. Some STs  become just {\em minimal} Einstein-scalar  (MES) theories, notable examples are  Brans-Dicke and  $R\phi^2$.  ST theories with {\em derivative}  coupling can also be reduced to an Einstein frame   by disformal transformations, but, as a rule, their scalar sector will contain higher derivative terms. Here we draw attention to a new Palatini kinetically coupled theory which can be reduced to pure MES  by an invertible disformal transformation. This theory can then be converted into the Jordan frame of another ST,  $R\phi^2$, which also admits an invertible transformation to MES. Two theories, each of which is dual to MES, will be {\em  sequentially  dual} to each other and can be considered as two different Jordan frames of the same theory. Both chosen theories   violate null energy condition. Transforming the same singular MES solutions, we find  the desingularization signs in both Jordan frames,  but these are more pronouned in the framework of kinetic theory, leading, in the cosmological case, to Genesis-type behavior.  
\end{abstract}  
\maketitle

\section{Introduction.}
Scalar-tensor theories with non-minimal coupling of a scalar to curvature have a long history, but  still remain the first choice in  search of a modified theory of gravity which could naturally explain inflation, dark energy and (possibly) dark matter,  for recent reviews see \cite{Capozziello:2011et,Nojiri:2017ncd,Heisenberg:2018vsk,Langlois:2018dxi,Barack:2018yly,Kobayashi:2019hrl}. Their unusual properties are closely related to violation of the energy conditions. Recall, that the strong energy condition  within the Friedmann-Robertson-Walker (FRW) cosmology, $\epsilon +3p\geq 0$,  implies that the universe is always slows down. This condition can be violated already by the minimally coupled massive scalar field with a potential, while the null energy condition (NEC), $\epsilon+p\ge 0$, is more robust; it is violated only in non-minimal STs, such as conformally coupled scalar-tensor theory \cite{Flanagan:1996gw,Mandal2019} or derivatively coupled theories: Horndeski, and more general degenerate higher order scalar-tensor theories (DHOST) \cite{Rubakov:2014jja}.
Violation of NEC drastically changes behavior of metrics near static or cosmological singularities both for non-derivative and derivative couplings, so it is interesting to study them in parallel.

A notable example of non-derivative ST violating NEC is the $\xi R \phi^2$ theory, which is conformal (in four dimensions) for $\xi=1/6$. Since the non-minimal term contains second derivatives of the metric,  its variation  gives rise to a stress tensor   first found by Chernikov and Tagirov \cite{Chernikov:1968zm}, then rediscovered in the QFT context,  as an ``impoved energy momentum tensor'' for the scalar field Callan: \cite{Callan:1970ze}, and further discussed by Parker \cite{Parker:1973kx} in curved spacetime (for some later discusstion see, e.g., \cite{Bellucci:2001cc}.  That this tensor violates various energy conditions was noted long ago by Beckenstein, who demonstrated  possibility of avoiding the cosmological singularity  \cite{Bekenstein:1975ww}, later work in this directions included \cite{Bayin:1994nz,Saa:2000ik,Gunzig:2000kk,Abramo:2002rn}. 

This theory also attracted attention in connection with inflation. For $\xi\neq 1/6$, the  $\xi R \phi^2$ theory  is no longer conformal. But more recently it was discovered that, in the case of large negative $\xi$, it may be useful for inflation. In fact, earlier attempts to associate inflation with the only physically known scalar field, Higgs,  were not successful for minimally coupled Higgs,   since the mass needed to accomodate   the observed density perturbations had to be of the order $10^{13}$ GeV and the quartic potential coefficient to be very small, $\lambda\sim 10^{-13}$. Situation could be improved by the non-minimal coupling  of the  type $\xi R \phi^2$  \cite{CervantesCota:1995tz}. With large negative $\xi$, the  tuning of the Higgs mass could be diminished \cite{ Bezrukov:2007ep}, but at the price of an unnaturally large $\xi$.
 Somewhat better was the situation with perturbations of the $\xi$-Higgs inflation in the Palatini approach to the same theory \cite{Bauer:2008zj}. Still there was a problem with unitarity  for quantized perturbations \cite{Burgess:2009ea}. To cure this, the derivatively coupled ST was suggested (new Higgs inflation)  \cite{Germani:2010gm}.  Combination of $\xi R \phi^2$ and derivative ST helped further improve the model \cite{Sato:2017qau}. Thus, derivative coupling turned out to be phenomenologically attractive. Moreover, it is well-known that derivative coupling can provide inflationary attractors without potential  terms at all \cite{Amendola:1993uh,Capozziello:1999uwa,Capozziello:1999xt,Sushkov:2009hk,Granda:2010hb}. 
 
Extremely popular became derivatively coupled STs after discovery of the ghost-free massive gravity and Galileon theories. This led to  Horndeski \cite{Horndeski} class, rediscovered as generalized Galileons \cite{Deffayet,Charmousis:2011bf}, beyond-Horndeski \cite{Zumalacarregui:2013pma,Gleyzes:2014qga} and derivative higher order STs (DHOST) theories including the whole set \cite{Langlois:2015cwa,Langlois:2018dxi}. Initially they were proposed in the metric formalism, but later also considered in Palatini \cite{Sotiriou:2005hu,Olmo:2009xy,Harko:2010hw,Olmo:2011uz,Luo:2014eda,Gumjudpai:2016ioy,Davydov:2017kxz,Galtsov:2018xuc,Jinno:2019und,Helpin:2019kcq} and hybrid \cite{Capozziello:2015lza,Olmo:2009xy}.

 Proliferation of derivatively coupled theories led to attempts to describe  general properties of the ST landscape \cite{Langlois:2018dxi,Quiros:2019gbt}. 
 An important tool for this is provided by {\em disformal dualities}. Introduced by Bekenstein \cite{Bekenstein:1992pj} on the basis of Finsler geometry as generalization of conformal transformations, they reappeared in derivatively coupled ST theories as   relations between different frames  \cite{Bettoni:2013diz,Domenech:2015hka,Sakstein:2015jca,Achour:2016rkg,BenAchour:2019fdf}. They can be used to obtain new Lagrangians, or as solution generation tools   \cite{Galtsov:2018xuc,BenAchour:2019fdf}, they also naturally arise in Palatini versions of STs \cite{Luo:2014eda,Galtsov:2018xuc,Helpin:2019kcq}. Special class constitute {\em invertible} disformal transformations: these do not change the number of degrees of freedom \cite{Exirifard:2007da,Deruelle:2014zza,Tsujikawa:2014uza,Arroja:2015wpa,Domenech:2015tca,deRham:2016wji,Takahashi:2017zgr,Babichev:2019twf}, so two STs related by an invertible disformal transformation  mathematically are equivalent.  

Here we want to draw attention to  group property of invertible transformations both conformal and disformal: two  successive transformations generate another invertivle transformation (up to subtleties with their domains). Consequently, two different ST theories, each of which can be transformed to an Einstein frame, in which the metric sector is described by the Einstein-Hilbert action, and  scalar sector is the same, will be  sequentially  dual to each other. If the scalar sector in the Einstein frame is described by the equations of the  second order, both dual STs will be free from Ostrogradsky instabilities. Of particular interest  is the class of STs which are invertibly reduced in their Einstein frames just to {\em minimal} Einstein-scalar theory (MES). Then you can use frame transformations as solution generating technique to explore new theories in situations which are considered problematic in the General theory of relativity, especially near singularities.
 
Recently, a new type of behavior has been discovered in cosmological solutions of STs with higher derivatives, such as Galileon theories \cite{Creminelli:2010ba} and DHOST \cite{Mironov:2019qjt}. The univers starts (or passes through the  previous evolution) from Minkowskyspace and demonstrates a sharp violation of NEC,  implying that the Hubble parameter satisfies the condition ${\dot H}\gg H^2$. In this case, the usual inflation scenario can be replaced by the alternative scenario called {\em Genesis}  \cite{Creminelli:2010ba}. It woud be interesting to know whether this behavior occurs in more familiar STs including non-derivative ones. Here we address this question using exact solutions which can be generated in the class of MES-dual theories. Modification of Penrose-Hawking  singularity theorems with weakened energy conditions was recently discussed in \cite{Fewster:2010gm,Brown:2018hym,Fewster:2019bjg}. 
 
Consider two different STs which reduce to MES in their respective Einstein frame and which, therefore, are sequentially dual.  Both Brans-Dicke and $\xi R \phi^2$ theories, as well some other  STs non-minimally coupled to scalar without derivatives, share this property, and the transition to their Einstein frame is done through invertible conformal transformations. Using any MES solution, it is possible to generate solutions of these two non-minimal STs  in their respective Jordan frames. Moreover, if the transformations between frames are invertible, one can start with a known Jordan frame solution of one ST, convert it to the Einstein frame, and then convert again into the Jordan frame  of another ST.  To find such dual theories on the set of derivatively coupled STs is a non-trivial task. Here we discuss one such theory which belongs to kinetically coupled class and which does not belong to the Horndeski class in the metric formalism (neither to DHOST).
 
 Desingularization in  $\xi R \phi^2$-theory is well-known.   Bekenstein found  the transformation to MES \cite{Bekenstein:1974sf} and applied it to transform the  Ficher-Janis-Newman-Winicour (FJNW) static spherically symmetric solution of a minimal theory with the singular ``would be'' horizon obtaining (with added Maxwell field) an asymptotically flat black hole found in 1970 by Bocharova, Bronnikov and Melnikov \cite{Bocharova:1970skc,Bekenstein:1975ts}. Bekenstein's duality was independently rediscovered and later discussed by many people \cite{Deser:1983rq,Futamase:1987ua,Schmidt:1988xi,Fakir:1990eg,Makino:1991sg}. Conversion to an Einstein frame (but not to MES) was found for non-minimal models including arbitrary functions $F(\phi)R$ and $F(R,\phi)$ \cite{Maeda:1988ab}, including the cosmological constant \cite{Galtsov:1992be} or   potentials  \cite{Abreu:1994fd} in the MES frame, in  higher dimensions \cite{Xanthopoulos:1992fm}. Later, a Palatini version of this theory was discussed, for relationship to the metric approach and references, see \cite{Bauer:2008zj}.

At the same time,  physical nonequivalence of the Jordan frame and the Einstein frame was  subject of long discussion, for a review of papers prior to 1994 see \cite{Magnano:1993bd,Abreu:1994fd,Wands:1993uu}, for more recent aspects and references see \cite{Faraoni:1996rf,Faraoni:1998qx,Faraoni:1999hp,Flanagan:2004bz,Sotiriou:2007zu}.  
As a rule,  two dual forms of scalar-tensor theory differ significantly when matter terms are added to them.  This actually determines which frame has to be considered as physical one. Another aspect of (non)equivalence is related to issues of  stability and   the quantum-level properties, tihis also remain the subject of discussion \cite{Sk.:2017fac}.
Here we will explore the difference of two Jordan frames of sequentially dual theories near the MES-frame singularities revealing that derivative coupling ensures stronger violation of NEC, than conformal coupling.  Namely, the static scalar MES singularity becomes a horizon in the $R\phi^2$ Jordan frame, but demonstrates regular behavior without a horizon in the new kinetic theory frame. The cosmological MES singularity becomes just the start of the universe from Minkowskyspacetime with the subsequent decelerating expansion within the non-derivative ST, but in the kinetic theory it exhibits  sharp violation of NEC at the very beginning of expansion generating the Genesis-type behavior.

The plan of the paper is as follows. In Section II we revisit the non-derivative $\xi R \phi^2$ discussing transformations to the Einstein frame,  NEC violation and other aspects. In Section III we consider the two-coupling derivative theory, which for some particular ratio of the couplings reduces to Horndeski class in the metric approach. We then adopt Palatini formulation, showing that the theory  is ghost-free for arbitrary couplings while for another ratio of two couplings the theory it is disformally dual to MES and, therefeore, sequentially  dual to the theory $\xi R \phi^2$. In Section IV we use dualities as a generating technique to construct Jordan frame duals for the static FJNW solution and the stiff-matter FRW cosmology in the Jordan frames of both theories, comparing their desingulariziation features. The results are summarized in  Section V.
 
\section{Non-derivative theory $\xi R \phi^2$ }
 
For the reader's convenience, we briefly review the main features of this theory, which is one of the oldest ST  with non-minimal non-derivative coupling \cite{Bekenstein:1974sf,Deser:1983rq,Futamase:1987ua,Schmidt:1988xi,Fakir:1990eg,Makino:1991sg}:
 \be 
\label{SC}
  S=  \int d^4x \sqrt{-g}\left(R  -g^{\mu\nu} \partial_{\mu}\phi
  \partial_{\nu}\phi - \xi R \phi^2 - 2V(\phi)
   \right), \ee
where we set $8\pi G_N=1$.
Variation of this  action
with respect to the metric and the scalar field   gives the
  Euler-Lagrange equations: 
 \be \label{eq}   
         G_{\mu\nu}  = 
T^{\phi}_{\mu\nu},  \qquad\qquad
\Box \phi  -\xi R \phi    =    0,  
       \ee
where the stress energy tensor is
\begin{equation}  \label{c}
 T^{\phi}_{\mu\nu} =
\partial_{\mu} \phi
\partial_{\nu} \phi -\frac{1}{2} g_{\mu\nu} g^{\alpha \beta}
\partial_{\alpha} \phi \partial_{\beta}
\phi  -g_{\mu\nu}V+\xi \left[g_{\mu\nu}  \Box -\nabla_{\mu}\nabla_{\nu}
 + G_{\mu\nu} \right] \phi^2 .
 \end{equation}
 
The Weyl transformation $\phi \to \Omega^{-1} \phi$,
 $g_{\mu\nu} \to \Omega^{ 2}
g_{\mu\nu}$,  leaves the Eqs. (\ref{eq})  invariant if $\xi=1/6$. In addtion,
$T^{\phi}_{\mu\nu} \to \Omega^{-2} T^{\phi}_{\mu\nu}$, if   $V=0$.  
Then the trace of $T^{\phi}_{\mu\nu}$  vanishes on shell \cite{Chernikov:1968zm,Callan:1970ze,Parker:1973kx}:
\begin{equation} \label{tracefi}
 g^{\mu \nu} T^{\phi}_{\mu\nu} = \phi \left( \Box \phi -
\frac{R}{6} \phi \right)=0,
\end{equation} 
and $R=0$ as expected for a conformal field, and so $\Box\phi=0$ on shell.

Attributing the Einstein tensor term in (\ref{c}) to the left hand side of the Einstein equation, we obtain the effective stress tensor:
\begin{equation}  \label{s}
G_{\mu\nu}=T^{\rm eff}_{\mu\nu}=(1-\xi\phi^2)^{-1}\left[\partial_{\mu} \phi
\partial_{\nu} \phi -\frac{1}{2} g_{\mu\nu} g^{\alpha \beta}
\partial_{\alpha} \phi \partial_{\beta}
\phi  -g_{\mu\nu}V+\xi \left(g_{\mu\nu}  \Box -\nabla_{\mu}\nabla_{\nu}\right)\phi^2 
  \right] 
 \end{equation}
 \subsection{Einstein frame}
 To pass to Einstein frame we recale the metric \cite{Bekenstein:1974sf, Futamase:1987ua}
 \be\lb{con}
 \X_{\mu\nu}=\Omega^2 g_{\mu\nu},\qquad \Omega^2=|1-\xi \phi^2|,
 \ee
 arriving at the following action
 \be 
\label{Shat}
  S=  \int d^4x \sqrt{-\X}\left(\R  -F^2(\phi)\X^{\mu\nu} \partial_{\mu}\phi
  \partial_{\nu}\phi - \V (\phi)
   \right), \ee
 where $ \R$ is the Ricci scalar of the new metric,  \be\label{Vhat}
    \V=\frac{V}{(1-\xi\phi^2)^2} ,\qquad  F^2=\frac{1-\xi(1-6\xi)\phi^2}{(1-\xi\phi^2)^2}, \ee
 are the new potential and the kinetic prefactor.
 
 To put the kinetic term into the standard form one has to pass to a new scalar field $\tph$, related to $\phi$ via
 \be\lb{phph}
 \frac{d \tph}{d \phi}=F(\phi). \ee
This redifinition results in the Einstein frame action 
 \be 
\label{EM}
  S_{\rm E}=  \int{d^4x \sqrt{-\X}\left(\R  -\X^{\mu\nu} \partial_{\mu}\tph
  \partial_{\nu}\tph  -\V (\tph) 
   \right),} \ee
  where the potential has to be expressed through the new scalar field. The Eq. (\ref{phph}) can  be integrated explicitly as follows \cite{Futamase:1987ua}:
  \be \tph=
   \begin{cases}
  \sqrt{\nu/\xi}\,\arcsin(\sqrt{\nu\xi}\,\phi)+\sqrt{3/2}\,\ln\Big|\frac{\sqrt{6}\xi\phi+\sqrt{1-\nu\xi\phi^2}}{\sqrt{6}\xi\phi-\sqrt{1-\nu\xi\phi^2}}\Big| ,  & \xi<1/6,\\
 \sqrt{3/2}\, \ln\Big|\frac{1+\sqrt{6}\xi\phi}{1-\sqrt{6}\xi\phi} \Big|, & \xi=1/6,\\
   \sqrt{|\nu|/\xi}\,{\rm arcsinh}(\sqrt{|\nu|\xi}\,\phi)+\sqrt{3/2}\,\ln\Big|\frac{\sqrt{6}\xi\phi+\sqrt{1-|\nu|\xi\phi^2}}{\sqrt{6}\xi\phi-\sqrt{1-|\nu|\xi\phi^2}}\Big| ,  & \xi>1/6,
 \end{cases}
  \ee 
  where $\nu=1-6\xi$.  

For $\xi=1/6, V=0$ these transformations reduces to the original form of conformal transformation found by Bekenstein \cite{Bekenstein:1974sf} and suggested as generating technique to construct solutions $R\phi^/6$ theory   from the solutions of MES:   from any solution $\X_{\mu\nu},\,\tph$ of the theory,
  \be 
\label{EM}
  S=  \int{d^4x \sqrt{-\X}\left(\R  -\X^{\mu\nu} \partial_{\mu}\tph
  \partial_{\nu}\tph  
   \right),} \ee
 a solution $g_{\mu\nu},\phi$ to the theory  $R\phi^/6$ theory is obtained via the transformation
 \be
 g_{\mu\nu}= (1-\phi^2/6)^{-1}\X_{\mu\nu} ,\qquad \phi=\sqrt{6}\tanh( \tph/\sqrt{6}) .
 \ee
This transformation is invertible, provided the value $\phi^2=6$ is not reached, an inverse map being
  \be\lb{invB}
 \X_{\mu\nu}=\cosh^2 ( \phi/\sqrt{6})\;g_{\mu\nu} ,\qquad \tph=\sqrt{6}(\tanh)^{-1}( \phi/\sqrt{6}).
 \ee
Maeda \cite{Maeda:1988ab}) had shown that a more general theory with the non-minimal functional coupling $F(R,\,\phi$ can be reduced to the Einstein-Hilbert term plus scalar fields (but not MES).

 \subsection{Generating Mexican hat potential}
Now let's
 start with the MES theory with the cosmological constant: 
  \be 
\label{EML}
  S_{\rm E}=  \int{d^4x \sqrt{-\X}\left(\R-2\Lambda  -\X^{\mu\nu} \partial_{\mu}\tph
  \partial_{\nu}\tph  
   \right),} \ee
   and apply the inverse duality transformations (\ref{invB}). The cosmological term then generats in the Jordan frame action  a  potential term \cite{Galtsov:1992be}:
  \be 
\label{SCH}
  S=  \int d^4x \sqrt{-g}\left(R  -g^{\mu\nu} \partial_{\mu}\phi
  \partial_{\nu}\phi -2V-\frac16 R \phi^2 
   \right), \ee
which has a Mexican hat shape
 \be
 V=\frac{\lambda}4 (\phi^2-v^2)^2, 
 \ee
 where in dimensionful units
 \be
 \lambda=\frac{8\pi G_N\Lambda}{9},\qquad v^2=\frac{3}{4\pi G_N},
 \ee
 and $G_N$ is the Newton constant.
  Note that the vacuum expectation value $v$ of  Higgs  is not a free parameter, but up to a factor is equal to the Planck's mass. In particular, one can not set $V=0$, so the resulting theory is not conformal.  The case of more general potentials in MES-frame was considered in \cite{Abreu:1994fd}.
 \subsection{Violation of NEC}
 The null energy condition for the effective stress-tensor reads
 \be
 T^{\rm eff}_{\mu\nu}l^\mu l^\nu\geq 0,\qquad l^\mu l_\nu=0,
 \ee
 for any null vector $l^\mu$. Substituting (\ref{s}), one obtains \cite{Mandal2019}:
 \be
   (1-\xi\phi^2)^{-1}\left[(\phi')^2-\xi(\phi^2)''\right]\geq 0,\ee
 where $\phi_\mu=\partial_\mu\phi,$ and the prime operation is defined as $ \phi'=l^\mu\nabla_\mu \phi.$
Therefore, for $\xi < 0$ , any local maximum of $\phi^2$
 violates NEC,
similarly for $\xi > 0$ , any local minimum of $\phi^2$  with $\xi\phi^2<1$ and
any local maximum of $\phi^2$  with $\xi\phi^2>1$ violate NEC.

 \subsection{Palatini}
 In the Palatini (or metric-affine) version \cite{Bauer:2008zj}, connection is treated as independent field which has to be fixed by varying the action $S_P(\hGamma, g)$:
  \be 
\label{SCP}
  S_P=  \int d^4x \sqrt{-g}\left(\R_{\mu\nu}(\hGamma) g^{\mu\nu}(1-\xi\phi^2)-g^{\mu\nu} \partial_{\mu}\phi
  \partial_{\nu}\phi -   2V(\phi)
   \right). \ee
Generically, independent variation of the connection generates non-metricity and torsion. In this case the Ricci tensor is not symmetric. However, the action  (\ref{SCP}) includes only symmetric part of it. As a result, it is invariant under {\em projective transformation} of the connection (for a recent discussion see  \cite{Shimada:2018lnm}
\be\lb{proj}
\Gamma_{\mu\nu}^\lambda\to \Gamma_{\mu\nu}+A_\mu\delta^\lambda_\nu,
\ee
in which case torsion can be consistently set to zero \cite{Bernal:2016lhq,Afonso:2017bxr}.  
  Then the Ricci tensor $\R_{\mu\nu}(\hGamma)$   should be varied as
   \be\lb{var}
 \delta \R_{\mu\nu}=\hnabla_\lambda \delta\hGamma^\lambda_{\mu\nu}- \hnabla_\nu \delta\hGamma^\lambda_{\mu\lambda},
   \ee
   where the covariant derivative with respect to the Palatini connection is understood.
   Variation of (\ref{SCP}) with respect to $\hGamma$, after integration by parts, leads to the following equation
   \be
   \hnabla_\lambda\left[g^{\mu\nu}(1-\xi\phi^2)\sqrt{-g}\right]=0.
   \ee
  With the field redefinition (\ref{con}), one can rewrite this as
  \be
  \hnabla_\lambda(\X^{\mu\nu}\sqrt{-\X})=0,
  \ee
  showing that the Palatini connection is nothing but the Levi-Civita connection of the Einstein frame metric.
  
  Variation of (\ref{SCP}) with respect to metric $g_{\mu\nu}$ gives the Einstein equation which can be written in terms of the Einstein frame metric as follows
  \be
  \R_{\alpha\beta}\left(\delta^\alpha_\mu\delta^\beta_\nu-\frac12\X_{\mu\nu}\X^{\alpha\beta}\right)=\frac{\phi_\mu\phi_\nu}{1-\xi\phi^2}-\frac12\X_{\mu\nu}\left(\phi_\alpha\phi_\beta\X^{\alpha\beta}+2V\right). \ee
  
 \section{Derivative coupling}
 \subsection{The metric theory}
Consider the action with non-minimal coupling of the scalar filed to Ricci tensor and Ricci scalar defined by the Levi-Civita connection
\begin{equation} \label{dm}
  S=\int d^{4}x\sqrt{-g}\left[R-\left( g_{\mu\nu}+\kappa_{1} g_{\mu\nu}R+\kappa_{2}R_{\mu\nu}\right)\phi^{\mu}\phi^{\nu}-2V(\phi)\right],
\end{equation}
where $\phi^\mu=\phi_nu g^{\mu\nu}$ and two coupling constants have dimension of inversed mass square. The Ricci scalar is defined though the Levi-Civita connection of the metric $g_{\mu\nu}$, its variation is given by
\begin{equation}
  \delta R_{\mu\nu}=\nabla^{\lambda}\nabla_{(\mu}\delta g_{\nu)\lambda}-\frac{1}{2}\Box \delta g_{\mu\nu}-\frac{1}{2}g^{\lambda\rho}\nabla_{\mu}\nabla_{\nu}\delta g_{\lambda\rho}.
\end{equation}
Applying this to (\ref{dm}) and commuting some covariant derivatives one obtains the equation
\begin{equation}
  G_{\mu\nu}= T_{\mu\nu}+\kappa_{1}\Theta_{\mu\nu}^1+\kappa_{2}\Theta_{\mu\nu}^2,
\end{equation}
where the first terms is the minimal energy-momentum tensor, 
$
 T_{\mu\nu}= \phi_{\mu}\phi_{\nu}-g_{\mu\nu}\left(\phi_\lambda\phi^\lambda/2-V(\phi)\right),
$
while the other terms correspond to separate contributions of two non-mininmal couplings
\begin{align}
  \Theta_{\mu\nu}^{1}&=\phi_\mu\phi_\nu R -\phi_\lambda\phi^\lambda G_{\mu\nu}   +\left(g_{\mu\nu}\Box- \nabla_{\mu}\nabla_{\nu}\right)(\phi_\lambda\phi^\lambda), \\
  \Theta_{\mu\nu}^{2}&=2\phi^\alpha\phi_{(\mu}R_{\nu)\alpha} -\phi^\alpha\nabla_\alpha \phi_{\mu\nu}+g_{\mu\nu}\left(\phi^{\alpha\beta}\phi_{\alpha\beta}/2
+(\Box \phi)^2 /2 + \phi^\alpha\nabla_\alpha \Box\phi\right), 
\end{align}
where $\phi_{\alpha\beta}\nabla_\alpha \phi_{\beta}$ and $G_{\mu\nu}$ is the Einstein tensor.
Variation over $\phi$ gives the scalar equation 
\begin{equation}\lb{phim}
\Box\phi+\nabla_{\mu}\big[\nabla_{\nu}\phi(\kappa_{1}g^{\mu\nu}R+\kappa_{2}R^{\mu\nu})\big]=0.
\end{equation}
Obviously, for generic values of the coupling constants  $\kappa_{1}$ и $\kappa_{2}$ both the Einstein and the scalar equations contain higher derivatives of 
$\phi$.  Collecting the third derivative terms, we find: 
\be \lb{third}
 \Theta_{\mu\nu}^{3} =(\kappa_2+2\kappa_1) \left( g_{\mu\nu}  \phi^{\alpha} \nabla_{\alpha}\Box\phi -\phi^\alpha\phi_{\alpha\mu\nu}\right).
\ee
These terms vanish in the case
\be
\kappa_2+2\kappa_1=0
\ee
corresponding to the Einstein tensor in the Lagrangian (\ref{dm}). The Ricci-terms in the scalar equation combine in the Einstein tensor too, so, in view of the Bianchi identity $\nabla_\mu G^{\mu\nu}=0$, which holds in the metric theory, the Eq. (\ref{phim}) becomes the second order eqiation
\begin{equation}
  \big[g^{\mu\nu}+\kappa G^{\mu\nu}\big]\nabla_{\mu}\nabla_{\nu}\phi=0.
\end{equation}

\subsection{ Palatini  }
In the Palatini version the action will read 
\be\lb{kP}
 S =\int d^{4}x\sqrt{-g}   
\left [(\R_{\mu\nu}-\phi_\mu\phi_\nu) g^{\mu\nu}-\R_{\alpha\beta} \phi_\mu\phi_\nu (\kappa_1   g^{\alpha\beta}g^{\mu\nu}  +\kappa_{2}   g^{\alpha\mu}g^{\beta\nu})\right].
\ee
Similarly to the conformally coupled theory,  this action includes only symmetric part of the Ricci tensor, and it is projective invariant under (\ref{proj}). We therefore set torsion to zero and make variation with respect to connection according to (\ref{var}).
This gives the following equation for an unknown connection:
\be \lb{eqZ}
 \hnabla_\lambda \left( \sqrt{-g}Z^{\mu\nu}\right) = 0,  \qquad Z^{\mu\nu}
  =\fL g^{\mu\nu} - \kappa_2\phi^\mu\phi^\nu, \qquad \phi^\mu=\phi_\alpha g^{\alpha\mu} \ee 
  where we have denoted
  \be \fL = (1 - \kappa_1 X), \qquad X= \phi_\alpha\phi_\beta g^{\alpha\beta}. 
\ee 
To solve the Eq. (\ref{eqZ}) with respect to $\hGamma$ we would like to cast it into the form $\hnabla_\la \X_{\mu\nu}=0$ for some second metric, or to some equivalent equation. Indeed, since $Z^{\mu\nu}\sqrt{-g}$ is the tensor density we will try to introduce such a metric via an identification 
\be\lb{ZX}
Z^{\mu\nu}\sqrt{-g}=\X^{\mu\nu}\sqrt{-\X},
\ee
so that the determinant would be of the same metric.
To proceed, we first construct the matrix $W_{\mu\nu}$, an inverse of the matrix $Z^{\mu\nu}$:
$$ W_{\mu\la}Z^{\la\nu}=\delta^\nu_\mu.$$  It can be obtained as linear combination of $g_{\mu\nu}$ and $\phi_\mu\phi_\nu$ as follows
 \be \lb{W}
W_{\mu\nu}=\fL^{-1} \left(g_{\mu\nu} +\kappa_2\fl^{-1}\phi_\mu\phi_\nu\right),
 \ee
 where $ \fl=1 -(\kappa_1+\kappa_2)X$.   To find the ratio of the determinants, we rewrite this in the form
 \be\lb{Wd}
 W_{\mu\nu}=\fL^{-1} g_{\mu\la}\left(\delta_\nu^\la+M_\nu^\la \right),\quad M_\nu^\la=\kappa_2\fL^{-1}\phi^\la\phi_\nu,
 \ee
where the matrix $M$ has the property $M^2\sim M$. For such matrices the determinant is given by
\be
\det(1+M)=1+{\rm tr} M.
\ee
Then from (\ref{Wd}) we obtain
\be
\det W=\fL^{-4} \det g \left( 1+\kappa_2 X/\fl\right)=\fL^{-3}\fl^{-1}\det g.
\ee
 Since the determinant of $Z^{\mu\nu}$ is inverse to $\det W$, we finally find from (\ref{ZX}) :
 \be\lb{det}
 \X=g\,\fl \fL^3,
 \ee
 and, using this, we obtain the second metric explicitely as
  \be\lb{dsf}
\X_{\mu\nu}=\sqrt{\fl\fL}\left(g_{\mu\nu}+\kappa_2 \fl^{-1} \phi_\mu \phi_\nu  \right).
 \ee
 Now the Eq. (\ref{eqZ}) becomes
 \be
 \hnabla(\X^{\mu\nu}\sqrt{-\X})=0,
 \ee
so the Palatini connection will be the Levi-Civita connection of the new metric:
 \be
 \hGamma^\lambda_{\mu\nu}= \X^{\la\tau}\left(\partial_\mu \X_{\la\nu}+ \partial_\mu\X_{\mu\la}-\partial_\la \X_{\mu\nu}\right)/2.
 \ee

Now we turn to other equations of motion. 
Variation of the action (\ref{kP}) with respect to the metric leads to the Einstein-Palatini equation 
\be \lb{PaE}
\fL\R_{\mu\nu}-\phi_\mu\phi_\nu(1+\kappa_1 \R)  -2\kappa_2 \R_{\alpha(\mu}\phi_{\nu)}\phi^\alpha- g_{\mu\nu}  L/2 = 0,
\ee
where the Lagrangian can be concisely presented as
\be\lb{LZ}
L=\R_{\mu\nu}Z^{\mu\nu}-\phi_\mu\phi_\nu g^{\mu\nu}
\ee
Finally,  a variation over $\phi$ gives rise to a scalar equation
\be\lb{sp}
\partial_\mu\left[ \sqrt{-g} \left( \phi^\mu+\kappa_1 \R\phi^\mu+\kappa_2
\R_{\alpha\beta}g^{\beta\mu}\phi^\alpha\right)\right]=0,
\ee
which, in principle,  could contain  higher-derivative terms. 

 \subsection{Einstein frame}
 So far we have obtained the second metric $\X_{\mu\nu}$ as an auxiliary one, needed to generate the Palatini connection. Note that it is related to the physical metric $g_{\mu\nu}$ by a disformal transformation (\ref{dsf}). The inverse  of $\X_{\mu\nu}$ can be read off from the Eq.(\ref{ZX}) with account for the ratio of determiants (\ref{det}):
 \be\lb{XX}
 \X^{\mu\nu}=\fl^{-1/2} \fL^{-1/2}\left( g^{\mu\nu}-\kappa_2 \phi^\mu\phi^\nu/\fL\right).
 \ee
 The functions $\lambda$ and $\Lambda$ depend on the initial metric through the norm of the gradient of the scalar field $X=\phi_\mu\phi_\nu g^{\mu\nu}$, so to invert the transformation one has to express $X$ through the norm with respect to the second metric ${\hat X}=\X^{\mu\nu}\phi_\mu\phi_\nu$. Contracting the Eq. (\ref{XX}) with $\phi_\mu\phi_\nu$ we obtain the equation
 \be\lb{Xeq}
 \hat X=X \fl^{1/2}\fL^{-3/2}.
 \ee
Clearly, we have to restrict physical domain by the conditions $\fl>0,\; \fL>0$. One must also avoid the critical point of the function ${\hat X}(X)$ where the derivative 
\be
\frac{\partial{\hat X}}{\partial X}=\frac{2-X(2\kappa_1+3\kappa_2)}{2\fl^{1/2}\fL^{5/2}}
\ee
is zero. This occurs at
\be
X=X_{\rm cr}=\frac2{2\kappa_1+3\kappa_2},
\ee
where the inverse derivative will diverge. But in the regions of monotonicity of ${\hat X}(X)$ the Eq. (\ref{Xeq}) is a cubuc equation obtained by squaring (\ref{Xeq})
\be\lb{Xla}
{\hat X}^2(1-\kappa_1 X)^3-X^2\left[1-(\kappa_1+\kappa_2)X\right]=0,
\ee
whose roots can be found explicitly (for more details see  \cite{Galtsov:2018xuc}), so
with such precautions, we can say that the transformation between two metrics is reversible.

Noticing the relation
\be
X\sqrt{-g}= {\hat X}\fl^{-1}\sqrt{-\X},
\ee
and  the representation (\ref{LZ}) of the Lagrangian,
it is now an easy task to express it entirely in terms of the second metric:
\be
  \sqrt{-g} L=\sqrt{-g}\left( \R_{\mu\nu} Z^{\mu\nu}-X\right)=\sqrt{-\X}\left( \R_{\mu\nu}  -\fl^{-1}\phi_\mu\phi_\nu\right) \X^{\mu\nu}.
  \ee
 We have obtained the Einstein-Hilbert term plus a modified scalar kinetic term without higher derivatives. In view of invertibility of the transformation to the Einstein frame, this means that the initial Palatini theory (\ref{kP}) is free of Ostrogradsky ghosts for general generic coupling constants $\kappa_1,\,\kappa_2$. Recall that in the metric formalism it belongs to Horndeski class only for $\kappa_2=-2\kappa_1$.

\subsection{New Palatini kinetic coupling}
 Now we see that in the Palatini formalism another particular relation, namely,
 \be\lb{kaka}
 \kappa_2=-\kappa_1=\kappa
 \ee
 defines an exceptionally simple derivetively couples  ST theory, 
 \be\lb{kP}
 S =\int d^{4}x\sqrt{-g}   
\left [(\R_{\mu\nu}-\phi_\mu\phi_\nu) g^{\mu\nu}-\kappa\R_{\alpha\beta} \phi_\mu\phi_\nu ( g^{\alpha\mu}g^{\beta\nu}- g^{\alpha\beta}g^{\mu\nu})\right],
\ee
in which case $\fl=1$ so it is disformally dual to MES is   in the Einstein frame  \cite{Galtsov:2018xuc}:
  \be\lb{MES}
S_E=\int \sqrt{-\X}\left[R_{\mu\nu}(\X)-\phi_\mu\phi_\nu\right]\X^{\mu\nu} d^4x.
 \ee
 In this dual theory the Einstein equation reads
 \be\lb{EX}
 R_{\mu\nu}=\phi_\mu\phi_\nu,
 \ee
 and the scalar obeys the covariant d'Alembert equation
\be\lb{sb}
{\hat\Box}\phi=0.
\ee
Note, that for the Einstein-Hilbert lagrangian both the metric and the Palatini variations lead to the same equations, therefore, one can replace the Palatini
Ricci scalar built with the Levi-Civita connection of the  Einstein frame metric, by the
 usual metric scalar curvature
 \be
 \X^{\mu\nu}\,\R_{\mu\nu}(\hGamma)= R(\X).
 \ee
 
 One can verify that Eqs. (\ref{PaE}) and (\ref{sp}) are satisfied by virtue of Eqs. (\ref{EX}) and (\ref{sb}). First, we obtain that Eq. (\ref{EX}) implies
$ L = 0, \, \R = \psi $, hence Eq. (\ref{PaE}) holds. Using then Eq. (\ref{EX}) in Eq. (\ref{sp}), we reduce the latter to (\ref{sb}).
 For this one-parametric family of Lagrangians (note that both signs of $\kappa$ are relevant, depending on whether the $\phi_mu$ is timelike or spacelike in the Einstein frame \cite{Galtsov:2018xuc}).
 
We will be interested in an inverse disformal transformation from Einstein metric $\X_{\mu\nu}$ to Jordan metric $g_{\mu\nu}$. For this, one has to express the factor $\fL$ through the Einstein-metric norm ${\hat X}=\phi_\mu\phi_\nu\X^{\mu\nu}$.
From the Eq. (\ref{Xla}) with account for (\ref{kaka}) one obtains the following cubic equation for $\sqrt{\fL}$:
\be
2z\left(\sqrt{\fL/3}\right)^3+\fL
-1=0,\qquad z=\frac{3\sqrt{3}}2\kappa_1\phi_\mu\phi_\nu\X^{\mu\nu},
\ee
 which has a real solution
 \be\lb{laf}
 \fL^{1/2}= \frac{\sqrt{3}}{2z}\begin{cases}
  2\cos\left(\frac13\arccos(2z^2-1)\right)-1, &\!\!z<1,\\
A^{1/3} +A^{-1/3}-1, & \!\!z>1,
 \end{cases}
 \ee
 where $A=2z\sqrt{z^2-1}+2z^2-1$. Then the Jordan metric will read:
 \be \lb{gg}
 g_{\mu\nu} = \X_{\mu\nu}\fL^{-1/2}+\kappa_1\phi_\mu\phi_\nu .\ee
 
 \section{Resolution of static singularities}
 \subsection{FJNW in the Einstein frame}
 Minimal scalar gravity (\ref{EM}) has a satic spherically symmetric solution, which was first found Fisher \cite{Fisher} and later rediscovered by many people including Janis, Newman and Winicour \cite{JNW}, nowadays mostly abbreviated as FJNW solution  
\begin{align}
&d\s^2=-\left(1-\frac{b}{r}\right)^{\gamma}dt^2+\left(1-\frac{b}{r}\right)^{-\gamma}dr^2+r^{2}\left(1-\frac{b}{r}\right)^{1-\gamma}(d\theta^2+\sin^2\theta d\varphi^2),\nonumber\\
& \tph=\frac{q}{b}\ln\left(1-\frac{b}{r}\right),
\end{align}
\noindent where $q$ is the scalar charge and $$\gamma=\left(1- \frac{2q^2}{b^2}\right)^{1/2} , \qquad    0<\gamma<1.$$  It is asymptotically flat and has a singularity at $r=b$.
\subsection{Conformal theory}
Consider the case $\gamma=1/2$ when all irrational powers are square roots. Then $q=b\sqrt{3/8}$  and Bekenstein's transformation reads
\begin{align}
	& \phi=\sqrt{6}\tanh(\tph/\sqrt{6})=\frac{\sqrt{1-b/r}-1}{\sqrt{1-b/r}+1},\\
	& ds^2=(1-\phi^2/6)^{-1} d\s^2.
\end{align}
Now perform the coordinate transformation
\be
1-{b}/{r}=\left(1-{b}/{(2\rho^2)}\right)^2.
\ee
 In terms of the new coordinates the solution takes the BBMB form
\begin{align}
	& \phi=\frac{\sqrt{6}m}{\rho-m},\quad m=\frac{b}4\\ 
	& ds^2=-\left(1-\frac{m}{\rho}\right)^{2} dt^2+\left(1-\frac{m}{\rho}\right)^{-2} d\rho^2+\rho^2d\Omega.
\end{align}
The metric conincides with the extremal Reissner-Nordstrom solution, while the  scalar field diverges on the horizion. As was shown by Bekenstein, the singulatity is unseen by a particle interacting with this scalar, so the solution as a whole can be regarded as a regular black hole. Thus a naked sungularity of MES solution was converted to a horizon in the frame of $R\phi2$ theory. But the singularity insode the horizon remained.

\subsection{New kinetic theory}
Now transform FJNW to the Jordan frame of the new kinetically coupled theory (\ref{kP}). In the static case, interesting solutions arise for  $\kappa_1=-\kappa_2>0$,  so here we denote $\kappa =\kappa_1 $ (or invert the sign of  $\kappa$ in (\ref{kP}) taking $\kappa$ positive again).  The disformal transformation (\ref{gg}) generates now the new metric aacording to the rules
\be
g_{tt}=\frac{\X_{tt}}{\fL^{1/2}},\qquad g_{rr}=\frac{\X_{rr}}{\fL^{3/2},}\qquad g_{\theta\theta}=\frac{\X_{r\theta,\theta}}{\fL^{1/2}},
\ee
where the factor $\fL$ is obtained from (\ref{gg}):
\be\lb{grr}
\fL^{-3/2} =\left(1-\frac{b}{r}\right)^\gamma  \left\{\frac{2x}{3\sqrt{3}} +\frac1{\sqrt{3}} 
 \begin{cases}
  2w\cos[\frac13\arccos(x/w)] ,&    x<w,\\
  w^{2/3}B+w^{4/3}B^{-1} , &  x>w,
 \end{cases}\right\}
\ee with
\be    B=\left(x+\sqrt{x^2-w^2}\right)^{1/3},\qquad x=\frac{3\sqrt{3} \kappa q^2}{2 r^2 (r-b)^2}.
\ee
For large $r$ the variable $x\sim 1/6^4$, so $\lambda=1+O(r^{-4})$ and the solution remains asymptotically flat:
\be
g_{tt}\sim -1+\frac{\gamma b}{r},\qquad g_{rr}\sim 1-\frac{\gamma b}{r}	,\qquad g_{\theta\theta}\sim r^2.
\ee
 Near the MES singularity $r=b$ one can expand in terms of $\xi=(r-b)/b$, denoting $\kappa q^2/b^4=\nu^3$:
\be
\nu^{-1} ds^2=-\xi^{2(2\gamma-1)/3} dt^2+(\nu dr/\xi)^2 dr^2 +b^2\xi^{(1-2\gamma)/3}(d\theta^2+\sin^2\theta d\varphi^2).
\ee
 In the case $\gamma=1/2$,  making the coordinate change $z=\mu b \ln\xi$ with $\infty<z<\infty$, one obtains
 \be
\nu^{-1} ds^2=- dt^2+ dz^2 +b^2 (d\theta^2+\sin^2\theta d\varphi^2).
\ee
 This metric represent the product of a two-dimensional Minkowsky space and a sphere. Note that the scalar field is not transformed and remains singular. But the disformal transformation appropriately subtracts divergence from the metric.  
 
\section{Cosmology}
   \subsection{MES cosmology with $\Lambda$}
Consider homogeneous and isotropic cosmologies in Einstein's theory minimally coupled to scalar in presence of cosmological constant. With the metric parametrization
\be
d\s^2=-\N^2 dt^2+\a^2 dl^2_k,\quad dl^2_k=d\chi^2+f_kd\Omega^2,
\ee 
where $k=-1,\,0,\,1$, with $f_1=\sin^2\chi,\, f_0=\chi^2,\,f_{-1}=\cosh^2\chi$ for spatially closed, flat and open universes respectively, and the functions $\N,\,\a$ depend only on $t$. (Our time coordinate and the three-space coordinates are dimensionless, while $\N,\,\a$ have dimension of kength.) We obtain the following relevant components of the Ricci tensor:
\begin{align}
	& \R_{tt}=\frac{3\dot{\N}\dot{\a}}{\N\a}- \frac{3\ddot{\a} }{ \a}, \\ 
	& \R_{\chi\chi}=\frac{\a\ddot{\a} }{ \N^2}-\frac{\a\dot{\N}\dot{\a}}{\N^3}+ \frac{2\dot{\a}^2 }{ \N^2}+2k .\lb{chch}
\end{align}
The Einstein equations read
\be
\R_{\mu\nu}=\Lambda \X_{\mu\nu}+\partial_\mu\tph \partial_\nu\tph.
\ee
 The component ($\chi\chi$) does not contain the scalar field and admits the first integral
 \be
 \frac{\a^4\dot{\a}^2 }{ \N^2}+k\a^4+\frac13\Lambda \a^6=a_0^4,
 \ee
 using which we find
 \begin{align}
	&\N^2=\frac{\a^4\dot{\a}^2} { a_0^4-k\a^4+\Lambda \a^6/3}, \\ 
	& \dot{\tph}^2=\frac{6a_0^4\dot{\a}^2 }{ \a^2\left( \beta^2-k\a^4+\Lambda \a^6/3\right)}. 
\end{align}
We still have freedom to fix the gauge, the convenient one being $\a=2a_0 t$. Then
\begin{align}
	& \N^2=\frac{(2a_0)^6 t^4 } { a_0^4 -kt^4+\Lambda t^6/3}, \\ 
	& \dot{\tph}^2=\frac{6a_0^4 }{ t^2\left( a_0^4-kt^4+\Lambda t^6/3\right)}. 
\end{align}

For more recent MES solutions which can be used as seed to probe non-minimal STs, see\cite{Astorino:2014mda,Sultana:2015lja,Banijamali:2019gry}.
 \subsection{Minkowsky start in $R\phi^2$} Performing Bekenstein's transformations in the case  $\Lambda=0,\, k=0$ one obtains the following exact cosmological solution of the theory (\ref{SCH}):
 \begin{align}
	  &\phi/\sqrt{6}=\tanh(\tph/\sqrt{6})= \frac{t^2 -1}{t^2 +1},\lb{Zreg1}\\
 & ds^2=(1-\phi^2/6)^{-1} d\s^2=\frac{ (t^2 +1 )^2}{4 t^2}d\s^2=
  (t ^2+1 )^2 \left[- (4a_0t)^2dt^2+a_0^2dl^2_0\right].\lb{Zreg2}
 \end{align}
 In terms of the synchronous time
 \be\lb{tt}
 \tau=a_0 t^2(t^2+2),\qquad {\rm or}\qquad t^2=\sqrt{1+\tau/a_0}-1,
 \ee
 we obtain
 \be
 ds^2=-d\tau^2+a^2 dl^2_0,\qquad a=a_0( t^2+1)=a_0\sqrt{1+\tau/a_0},	
 \ee
 Thus the univers starts from Minkowsky space. The Hubble parameter and its derivative are
\be
H=\frac1{a}\frac{da}{d\tau}=\frac1{2(a_0+\tau)},\qquad {\dot H}=\frac{dH}{d\tau}=-2 H^2.
 \ee
 The universe is always decelerating.
 
When $k=\pm1, \Lambda\neq 0$ the very beginning of the expansion is the same.
 
 \subsection{New kinetic theory: Genesis}
 Now transform the MES cosmological solution to the Jordan frame of the Palatini kinetically coupled theory (\ref{kP}). In this case the relevant sign of the coupling constant $\kappa$ is positive.  We will be interested by behavior of the scale factor near the singularity of the MES solution. Since in this case both the cosmological constant and curvature terms are negligible, we start with $k=0,\,\Lambda=0$ in synchronous gauge
\be\lb{frw}
d\s^2=\X_{\mu\nu} dx^\mu dx^\nu=-dt^2+{\hat a}^2\delta_{ij}dx^i dx^j,
\ee
where   
\be\lb{Ze}
{\hat a}=a_0 t^{1/3},\qquad \phi=\sqrt{ 2 }\ln t/\sqrt{3},
\ee
as was found by Zel'dovich in 1972 for the ``stiff-matter'' \cite{Zeldovich,Chavanis:2014lra}. Obviously, this metric is singular at $t=0$ and describes a decelerating expansion.

Now we transorm to the Jordan frame of the kinetic theory. From (\ref{gg}) we obtain  an  algebraic equation for $N$:
\be
\left(N-2z/(3\sqrt{3})\right)^3=N^2,\quad z=\kappa \sqrt{3}/t^2.
\ee
Its real solution is   smooth, although in terms of real functions it looks piecewise:
\be\lb{gtt}
N^2 = \frac{2z}{3\sqrt{3}}+\frac1{\sqrt{3}}
 \begin{cases}
  2\cos\left(\frac13\arccos(x)\right), &z<1,\\
A^{1/3} +A^{-1/3}, & z>1,
 \end{cases}
\ee
where $A=\left( z+\sqrt{z^2-1}\right)^{1/3}$.
For large $z$ (small $t$) one has:
\be\lb{sing}
N^2 =  {2z}/{3\sqrt{3}}+{(2z)^{\frac13}}/{\sqrt{3}}+\left(4/{z}\right)^{\frac13}/{(2\sqrt{3})}+...\,,
\ee
for small $z$ (large $t$),
\be\lb{as}
N^2 =1+ {z}/{\sqrt{3}}- {z^2}/{18}+...\;,
\ee
In terms of time this gives
\be
g_{tt}=(\alpha t)^{-2}\left(1+(\alpha t)^{4/3} \right),\qquad \alpha=\left(\frac3{2\kappa}\right)^{1/2}.
\ee
For the scale factor we obtain:
\be
  a^2={\hat a}^2 N^{2/3}.
\ee
Now need to go to the synchronous time $t\to\tau(t)$  solving the equation $
Ndt=d\tau.$
For small $t$, keeping the leading term in (\ref{sing}), one finds:
\be
dt/d\tau=H_0 \,t \,\Longrightarrow \, t={\rm e}^{H_0 \tau},\,H_0=\sqrt{ 3/{(2\kappa)}}.
\ee
  
\be
dt/d\tau=\alpha \,t \,\rightarrow \, t={\rm e}^{\alpha \tau},
\ee
so that $t\to 0$ corresponds $\tau\to -\infty$. 

Now compute the Hubble parameter differentiating with respect to synchronous time in the vicinity of $t=0$:
\be
H=\frac1{a}\frac{da}{dt}\frac{dt}{d\tau}=2\sqrt{\alpha} (\alpha t)^{4/3}.
\ee  
Its derivative reads
\be
{\dot H}=\frac{dH}{d\tau}=\frac{2\alpha}9(\alpha t)^{4/3},
\ee
and satifies condition of strong NEC violation: the ratio
\be
\frac{\dot H}{H^2}=\frac92(\alpha t)^{-4/3}=\frac9{2\alpha^{4/3}}{\rm e}^{-4 \alpha \tau/3},
\ee
 diverges exponentially in terms of the synchronous time as $\tau \to - \infty$. Such behavior is typical for Genesis scenario \cite{Creminelli:2010ba,Mironov:2019qjt}.
So, NEC violation is even more pronounced in the kinetic theory.
 
\section{Conclusions}
Our goal was to discuss sequential duaities in  non-minimal scalar-tensor theories which arise when two or more theories coincide in their respective Einstein frames into which they can be transformed using invertible mappings. 
Such dualities are especially useful if the Einstein frame theory is simply the minimally coupled Einstein-scalar theory.
By the group property of reversible mappings, two theories, each of which is dual to the MES, are dual to each other; therefore, in a sense, they can be considered as two Jordan systems of the same theory. It is not difficult to find such theories bewtween the subset of non-derivatively coupled STs. As an example of derivative coupling, we have chosen the recently proposed new Palatini kinetc theory. Transforming  static and cosmological solutiopn of MES into Jordan frames of these two theories we have found that the second one drastically change behavior of solutions near the singularity and in the cosmologal case leads to Genesis-like behavior.

Class of sequential dualities can be extended taking MES with potentials, which also allow for exact solutions. These will generate non-minimal STs which will be ghost-free as well, though presumably they will not have such a simple form in their Jordan frame as our example here.

  We realize, of course, that adding matter generically will destroy suquential dualities in STs, but still such property of their pure gravitational sectors (including scalar degree of freedom)  seems useful in understanding the landscape of STs.

 \noindent {\textbf{\textit {Acknowledgments.}}}
The author is grateful to G\'erard Cl\'ement  for careful reading of the manuscript and  valuable comments. He also  thanks Evgeny Babichev, Andrei Barvinsky, Salvatore Cappozziello, Jose Beltr\'an Jim\'enez,  Sergei Sushkov and Michael Volkov for useful discussions. The work was supported by the Russian Foundation for Basic Research grant 17-02-01299a. The
networking support by the COST Action CA16104 is acknowledged.

\end{document}